\documentclass[9pt,twocolumn,twoside]{osajnl}

\journal{ol} 

\setboolean{shortarticle}{true} 
\usepackage{color}

\title{Photonic Loschmidt echo in binary waveguide lattices}

\author[1,*]{S. Longhi}
\affil[1]{Dipartimento di Fisica, Politecnico di Milano and Istituto di Fotonica e Nanotecnologie del Consiglio Nazionale delle Ricerche, Piazza L. da Vinci 32, I-20133 Milano, Italy}
\affil[*]{Corresponding author: stefano.longhi@polimi.it}

\dates{Compiled \today}

\ociscodes{(130.3120)   Integrated optics devices;  (270.5290)   Photon statistics;  (000.1600) Classical and quantum physics;
(070.6760)   Talbot and self-imaging effects}

\doi{\url{http://dx.doi.org/10.1364/ol.XX.XXXXXX}}

\begin{abstract}
Time reversal is one of the most intriguing yet elusive wave phenomenon of major interest in different areas of classical and quantum physics. Time reversal requires in principle to flip the sign of the Hamiltonian of the system, leading to a revival of the initial state (Loschmidt echo). Here it is shown that Loschmidt echo of photons can be observed in an optical setting without resorting to reversal of the Hamiltonian. We consider photonic propagation in a binary waveguide lattice and show that, by exchanging the two sublattices after some propagation distance, a Loschmidt echo can be observed. Examples of Loschmidt echoes for single photon and NOON states are given in one- and two-dimensional waveguide lattices.
\end{abstract}

\setboolean{displaycopyright}{true}

\begin{document}

\maketitle
\thispagestyle{fancy}
\ifthenelse{\boolean{shortarticle}}{\abscontent}{}

Time reversal is one of the most intriguing yet elusive
 wave phenomenon that has attracted a great attention of physicists since more than one century.  
 In principle,  changing the sign of a time-symmetric Hamiltonian can time reverse the evolution of a classical or quantum system.
 This result seems at odd with the irreversible dynamics observed in nature, a contradiction which is at the heart of the so-called Loschmidt paradox \cite{r1,r1bis,r2}. 
  Several experimental demonstrations of time reversibility have been reported for
quantum dynamics or classical waves, including spin systems \cite{r3}, acoustic \cite{r4}, electromagnetic \cite{r5}, water waves \cite{r6}, and 
  atom optics systems \cite{r7,r8,r9,r10}, with a wealth of interesting technological applications. In optics, time reversal dynamics, like the one based on optical phase conjugation, provides a powerful tool to eliminate aberrations or scattering of optical waves in inhomogeneous or diffusive media \cite{r11,r12,r13}. A short overview of time reversal symmetry in classical and quantum optics  can be found in \cite{r14}.\\ 
  The forward-backward evolution of an isolated quantum system induced by changing the sign of the Hamiltonian yields a revival or echo effect, the so-called Loschmidt echo \cite{r1,r1bis,r2}. Let $ |\psi(0) \rangle$ be the state of the system at initial time $t=0$, which evolves to the state $|\psi(T) \rangle= \exp(-i \hat{H}_1 T) | \psi(0) \rangle$ at the time $t=T$ with the Hamiltonian $\hat{H}_1$. In the successive time interval $T$, the Hamiltonian of the system is changed into $\hat{H}_2=-\hat{H}_1+\hat{V}$, where $\hat{V}$ is generally a small perturbation that accounts for non-perfect reversal or system-environment interaction. The final state $|\psi(2T) \rangle=\exp(i \hat{H}_1T-i\hat{V}T) \exp(-i \hat{H}_1 T) | \psi(0) \rangle$ at time $t=2T$ reproduces the initial state $|\psi(0) \rangle$ with an accuracy which is measured by the fidelity $\mathcal{F}=| \langle \psi(2T) | \psi(0) \rangle |$, with $\mathcal{F} \leq 1$ and $\mathcal{F}=1$ for perfect reversal $\hat{V}=0$. In quantum systems with few degrees of freedom, today' s technological advances make it meaningful to address time reversal experiments. So far most experiments on time reversal require to flip the sign of the Hamiltonian, i.e. $\hat{H}_2 \simeq - \hat{H}_1$. 
 In this Letter we suggest Loschmidt echo of photons in optical waveguide lattices without resorting to Hamiltonian reversal. We consider light propagation in binary waveguide lattices, comprising two sublattices A and B,  and show that exchange of the two sublattices after some propagation distance results in approximate time reversal. The Loschmidt echo results from self-imaging of the waveguide lattice, which turns out to be robust against imperfections or disorder in the lattice coupling constants. For high photon number states, a rapid decay of Loschmidt echo and resolution of the Loschmidt paradox can be observed in such a photonic simulator of time reversal dynamics.\\ 
 Propagation of classical and non-classical light in waveguide lattices has received a great interest in the past two decades (see e.g. \cite{r15,r16,r17,r18,r19,r20} and references therein). Such systems enabled direct observation of optical analogues of many fundamental quantum mechanical effects, such as Bloch oscillations \cite{r21,r22,r22bis}, Anderson localization \cite{r23,r24}, quantum Zeno dynamics \cite{r25}, dynamical localization \cite{r26}, and many others. Waveguide lattices also provide a rather unique platform to realize quantum interference effects, quantum walks and other sophisticated quantum manipulations of light in a robust, decoherence-free and integrated  environment \cite{r18,r19,r20,r27,r27bis}. Self-imaging effects in one-band waveguide lattices, based on a sudden application of a phase gradient to the optical wave, were  proposed and demonstrated in a few recent works \cite{r28,r29,r30,r31}. Such a self-imaging effect, also referred to as diffraction management or perfect imaging, can be regarded as a kind of time reversal dynamics because application of a sudden phase gradient to the optical wave is equivalent to changing the sign of the Hamiltonian $\hat{H}_2 =-\hat{H}_1$ \cite{r9,r31}. Here we consider a two-band lattice model, i.e. a binary waveguide lattice \cite{r22bis,r32}, and show that self-imaging dynamics can be realized, at least approximately, by exchanging the two sublattices A and B, which manifestly violates the condition $\hat{H}_2 =-\hat{H}_1$. We focus our analysis to a one-dimensional binary waveguide array, however the results can be readily extended to a two-dimensional lattice.
 Photons propagating in a binary waveguide array with nearest-neighbor coupling is described by the Hamiltonian \cite{r18,r19,r27bis}
 \begin{equation}
 \hat{H}_1= \sum_n \left( \kappa_n \hat{a}^{\dag}_n \hat{a}_{n+1}+{\rm H.c.} \right)+\delta \sum_n (-1)^n  \hat{a}^{\dag}_n \hat{a}_n
 \end{equation}
 where $\hat{a}^{\dag}_n$ is the creation operator of photons in the mode of waveguide $n$, $\kappa_n$ is the coupling constant between nearest neighbor waveguides $n$ and $(n+1)$, and $2\delta$ is the propagation constant offset between modes in the two sublattices A and B. The sublattices A and B correspond to the waveguides with index $n$ being even or odd, respectively. Along with the Hamiltonian $\hat{H}_1$, we consider the Hamiltonian $\hat{H}_2$ which is obtained from $\hat{H}_1$ by reversing the sign of $\delta$. Note that $\hat{H}_2$ is not the time reversal of $\hat{H}_1$, which would require to reverse {\it both} signs of $\delta$ and $\kappa_n$. In practice, $\hat{H}_2$ is the Hamiltonian that describes photon propagation in a waveguide array where the sublattices A and B have been interchanged. The main result of our analysis is that, if we consider photon propagation in a sequence of two lattices of the same length $L=cT$, with the two sublattices A and B interchanged in the second lattice, then approximate time reversal dynamics is realized provided that the propagation constant detuning $\delta$ is much larger than the coupling constants $\kappa_n$. The reversal (echo) effect of an initial quantum state $| \psi(0) \rangle$ exciting the array is described by the fidelity
  \begin{equation}
  \mathcal{F}= \left| \langle \psi(0) | \exp(-i \hat{H}_2 L) \exp(-i \hat{H}_1 L) | \psi(0) \rangle  \right|
  \end{equation}
  To show the effective time reversal dynamics despite $\hat{H}_2 \neq - \hat{H}_1$, let us consider the Heisenberg equations for the creation operators, which read
  \begin{equation}
  -i \frac{d \hat{a}^{\dag}_n}{dz}= \kappa_n \hat{a}^{\dag}_{n+1}+\kappa_{n-1} \hat{a}^{\dag}_{n-1}+(-1)^n \delta \hat{a}^{\dag}_n.
  \end{equation}
 For $\delta \gg \kappa_n$, at leading order (rotating-wave approximation) the operators $\hat{a}^{\dag}_{2n}$ and $\hat{a}^{\dag}_{2n+1}$ at even and odd lattice sites, i.e. in the two sublattices A and B, are almost decoupled and oscillate rapidly according to the relations $\hat{a}^{\dag}_{2n}(z) \simeq \hat{a}^{\dag}_{2n}(0) \exp(i \delta z)$ and $\hat{a}^{\dag}_{2n+1}(z) \simeq \hat{a}^{\dag}_{2n+1}(0) \exp(- i \delta z)$. A first-order correction in the small parameter $\kappa_n / \delta$, i.e. beyond the rotating wave approximation, can be obtained by standard asymptotic methods, yielding the following evolution equations for the creation operators
  \begin{eqnarray}
 - i \frac{d \hat{a}^{\dag}_{2n}}{dz} &  \simeq &  \frac{\kappa_{2n} \kappa_{2n+1}} {2 \delta} \hat{a}^{\dag}_{2n+2}+ \frac{\kappa_{2n-1} \kappa_{2n-2}} {2 \delta} \hat{a}^{\dag}_{2n-2} \nonumber \\
  & + &   \left( \delta +\frac{\kappa_{2n}^2+\kappa_{2n-1}^2}{2 \delta} \right) \hat{a}^{\dag}_{2n}  \\
  - i \frac{d \hat{a}^{\dag}_{2n+1}}{dz} &  \simeq & - \frac{\kappa_{2n+1} \kappa_{2n+2}} {2 \delta} \hat{a}^{\dag}_{2n+3}-\frac{\kappa_{2n} \kappa_{2n-1}} {2 \delta} \hat{a}^{\dag}_{2n-1} \nonumber \\
   & - & \left( \delta+ \frac{\kappa_{2n}^2+\kappa_{2n+1}^2}{2 \delta}
   \right)\hat{a}^{\dag}_{2n+1}
    \end{eqnarray}
  which can be derived from the effective Hamiltonian 
  \begin{eqnarray}
  \hat{H}_{1}^{(eff)} & = & \sum_n \left( \frac{\kappa_{2n} \kappa_{2n+1}}{2 \delta} \hat{a}^{\dag}_{2n} \hat{a}_{2n+2}
  -\frac{\kappa_{2n+1} \kappa_{2n+2}}{2 \delta} \hat{a}^{\dag}_{2n+1} \hat{a}_{2n+3} \right. \nonumber \\
  & + & \left.  {\rm H.c.} \right) + \sum_n \left( \delta +\frac{\kappa_{2n}^2+\kappa_{2n-1}^2}{2 \delta} \right) \hat{a}^{\dag}_{2n} \hat{a}_{2n} \\
  & - & \sum_n \left( \delta +\frac{\kappa_{2n+1}^2+\kappa_{2n}^2}{2 \delta} \right)  \hat{a}^{\dag}_{2n+1} \hat{a}_{2n+1} . \nonumber
  \end{eqnarray}
   \begin{figure}[htb]
\centerline{\includegraphics[width=8.4cm]{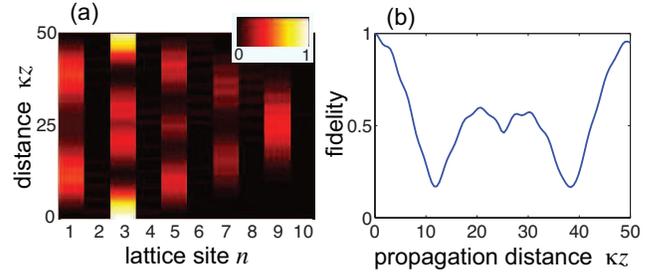}} \caption{ \small
(Color online) Evolution of (a) the mean photon number $\langle \hat{a}^{\dag}_n \hat{a}_n \rangle$ and (b) the fidelity $\mathcal{F}$ versus normalized propagation distance $\kappa z$ in a binary array made of $N=10$ waveguides with uniform hopping rate $\kappa$ and for $\delta / \kappa=5$. The array is initially excited with one photon state in waveguide $n=3$.}
\end{figure} 

 \begin{figure}[htb]
\centerline{\includegraphics[width=8.4cm]{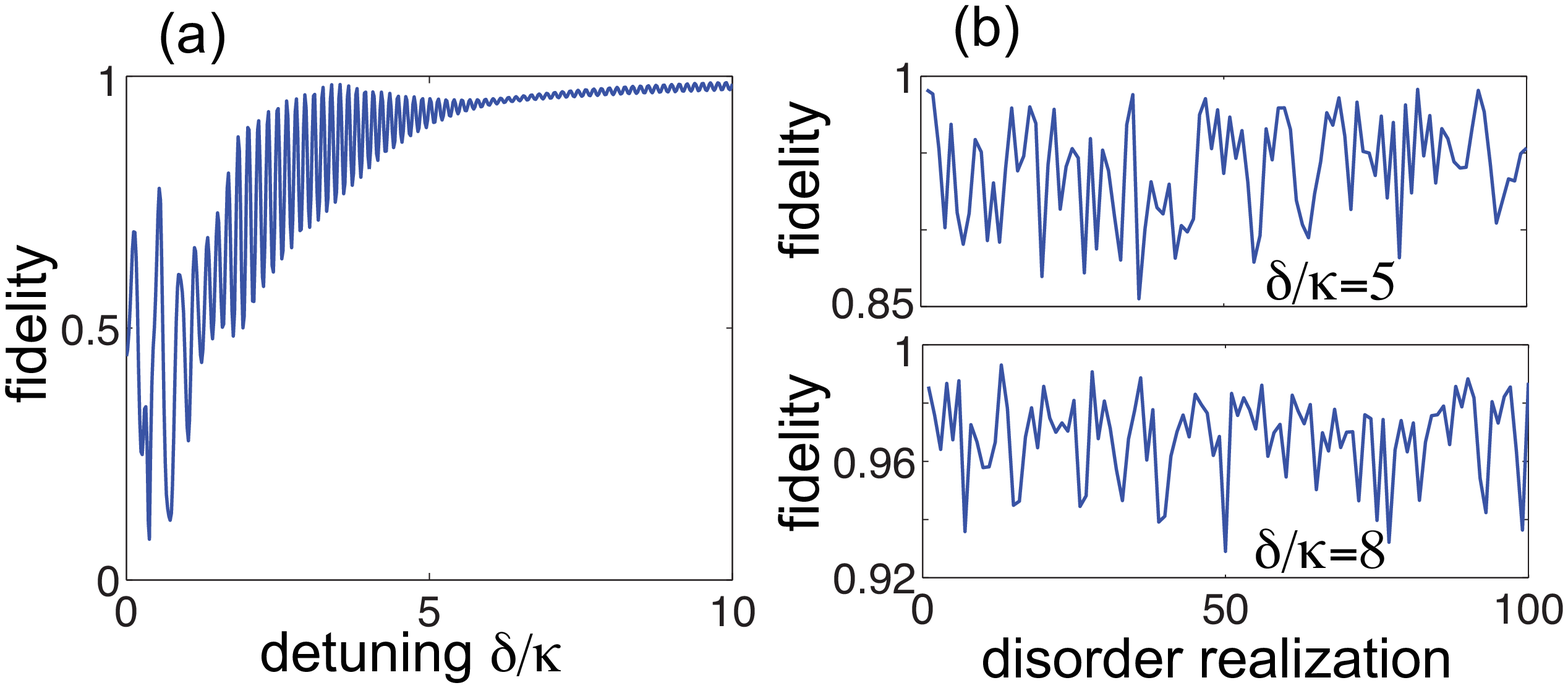}} \caption{ \small
(Color online) (a) Behavior of the fidelity $\mathcal{F}$ at the output plane versus the normalized detuning parameter $\delta / \kappa$ in a binary array made of $N=10$ waveguides with uniform hopping rate $\kappa$. (b) Behavior of the fidelity $\mathcal{F}$ at the output plane in 100 binary arrays with different realizations of disorder in coupling constants $\kappa_n$ and for two values of $\delta / \kappa$. The coupling constant between guides $n$ and $(n+1)$ is given by $\kappa_n=\kappa(1+ \sigma_n)$, where $\sigma_n$ is a random variable with uniform distribution in the range $(-0.2,0.2)$. In both (a) and (b) the array is initially excited with one photon state in waveguide $n=3$.}
\end{figure}   

 \begin{figure}[htb]
\centerline{\includegraphics[width=8.4cm]{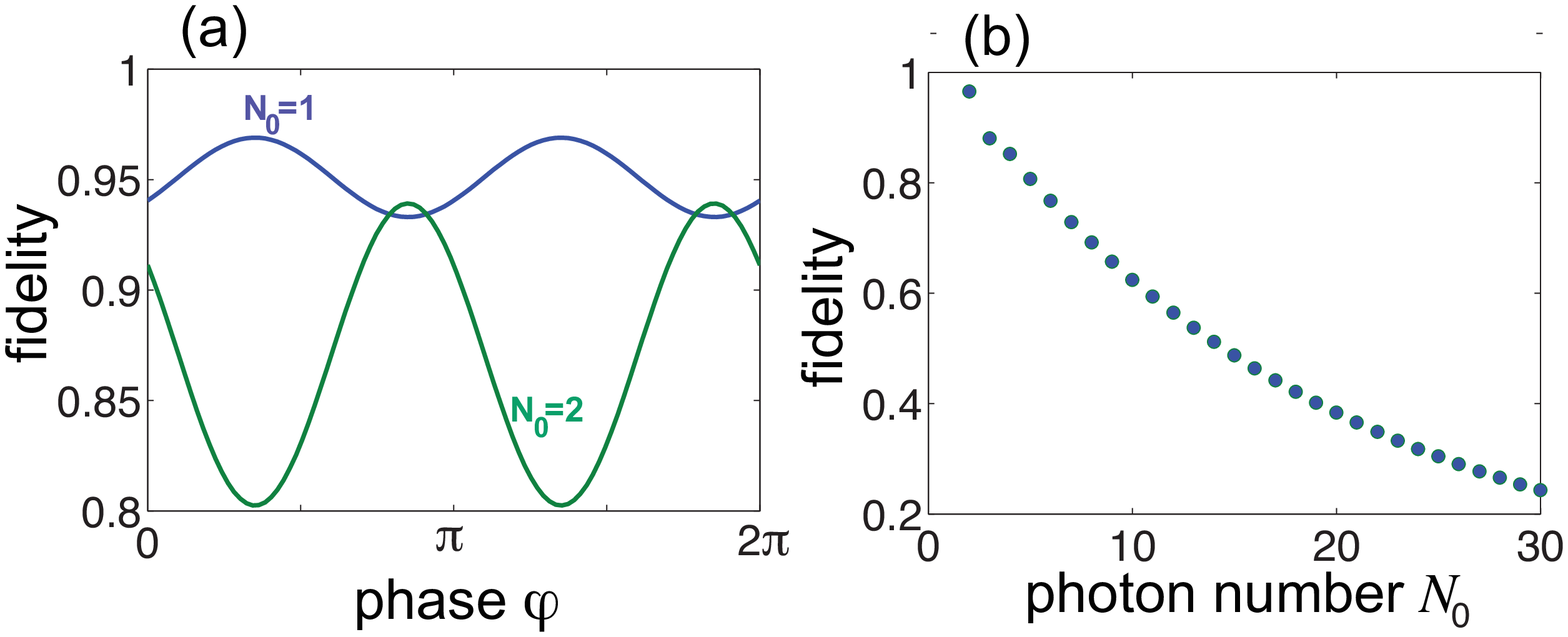}} \caption{ \small
(Color online) (a) Behavior of the fidelity $\mathcal{F}$ versus phase $\varphi$ at the output plane of the binary array of Fig.1 when it is excited by a  NOON state with $N_0=1$ (i.e. a Bell state) and $N_0=2$ photons at the two waveguides $n_1=1$ and $n_2=2$. (b) Behavior of the fidelity $\mathcal{F}$ for a NOON state excitation with $\varphi=0$ and for increasing number of photons $N_0$.}
\end{figure}

  Note that in such a limit the dynamics in the two sublattices is decoupled, their interaction being included in renormalized photonic couplings and propagation constant offsets. From a physical viewpoint, photon hopping between adjacent sites in the same sublattice is a second-order tunneling process \cite{rR1,rR2} mediated by the intermediate out-of-resonance sites of the other sublattice. The resulting hopping rates, defined in Eqs.(4) and (5), are inversely proportional to the propagation constant offset $\delta$; therefore they can be reversed by flipping the sign of $\delta$. Second-order tunneling also introduces effective shifts of site energies in the two sublattices, which are sensitive to the sign of $\delta$ as well \cite{rR1,rR2}. Therefore, by reversing the sign of $\delta$ from Eq.(6) it follows that  $\hat{H}_2^{(eff)}=-\hat{H}_1^{(eff)}$.  This means that time reversal can be approximately obtained, even thought the original (exact) Hamiltonian is not time reversed ($\hat{H}_2 \neq -\hat{H}_1$). For a homogeneous binary lattice, approximate time reversal can be physically explained as follows. In the homogeneous case $\kappa_n= \kappa$, the binary lattice sustains two mini bands with dispersion curves \cite{r32} $E_{\pm }(q)= \pm \delta \sqrt{1+(2 \kappa / \delta)^2 \cos^2 q}$, where $ -\pi/2 \leq q < \pi/2$ is the Bloch wave number. For $\delta \gg \kappa$, the Bloch eigenstates of the two mini bands correspond to occupation of either one of the two sublattices. Hence, by reversing the sing of $\delta$ the occupation in two mini bands is flipped. Since the dispersion curves in the two mini bands introduce opposite phase shifts at any wave number $q$, time reversal is obtained. The fidelity of the time reversal is expected to increase as the parameter $ \kappa / \delta$ is diminished. Note that such a result holds more generally for inhomogeneous hopping rates $\kappa_n$, i.e. time reversal is robust against lattice truncation and structural imperfections or disorder in the coupling constants. However, disorder in site energies, i.e. in $\delta$, can not be time reversed and can therefore prevent the observation of Loschmidt echo. Disruption of self-imaging by diagonal disorder is a detrimental effect which is common to other time-reversal methods \cite{r31}.\\ 
  We checked the validity of the theoretical analysis by numerical computation of the fidelity for different excitation conditions of the binary lattice. Equations (3) have been integrated from $z=0$ (input or excitation plane of the array) to $z=2L$ (output plane) assuming a sudden change of the sign of $\delta$ at $z=L$. The fidelity at plane $z$ for classical light, i.e. when the operators $\hat{a}^{\dag}_n$ in the equations are treated as $c$-numbers, is simply given by $\mathcal{F}(z)=(1/ \mathcal{N}) \left|  \sum_n  \alpha_n^*(z) \alpha_n(0) \right|$, where $\alpha_n(0)$ is the classical excitation amplitude of waveguide $n$ at $z=0$, $\alpha_n(z)$ is the propagated amplitude,  and $\mathcal{N}= \sum_n | \alpha_n|^2$ is the normalization constant. The fidelity for classical fields is equivalent to the one obtained when the array is excited by the single photon state $|\psi(0) \rangle=(1/ \sqrt{\mathcal{N}}) \sum_n \alpha_n(0)  \hat{a}^{\dag}_n |0 \rangle$.
  Figure 1 shows as an example the behavior of the fidelity $\mathcal{F}(z)=| \langle \psi(0) | \psi(z) \rangle|$ and of the mean photon number $\langle \hat{a}^{\dag}_n \hat{a}_n \rangle $ (classical light intensity) at various lattice sites versus propagation distance in an array made of $N=10$ waveguides with uniform couplings $\kappa_n=\kappa$, for $\delta / \kappa=5$ and for a total waveguide length $2L=50/ \kappa$. The array is excited  at the input plane by a single photon state in the guide $n=3$, i.e. $|\psi(0) \rangle=\hat{a}_3^{\dag} |0 \rangle$. The behavior of the fidelity at the output plane versus the ratio $\delta/ \kappa$ is shown in Fig.2(a), clearing indicating that $\mathcal{F}$ increases as the normalized detuning parameter $\delta / \kappa$ is increased. The effect of disorder in the coupling constants $\kappa_n$ on the fidelity $\mathcal{F}$ is shown in Fig.2(b) for two values of the detuning $\delta / \kappa$. The figure depicts the behavior of $\mathcal{F}$ for 100 different realization of disorder, where we assumed $\kappa_n =\kappa(1+\sigma_n)$ with $\sigma_n$ a random variable with uniform distribution in the range $(-0.2,0.2)$. The figure clearly demonstrates that the time reversal method is robust against disorder in coupling constants (off-diagonal disorder). However, like for other time reversal methods in a single-band lattice \cite{r31}, self-imaging and time reversal are degraded by on-diagonal disorder, i.e. disorder in the propagation constants of waveguides, which should be therefore avoided in the manufacturing of the waveguide lattice.\\ 
     \begin{figure*}[htb]
\centerline{\includegraphics[width=16.5cm]{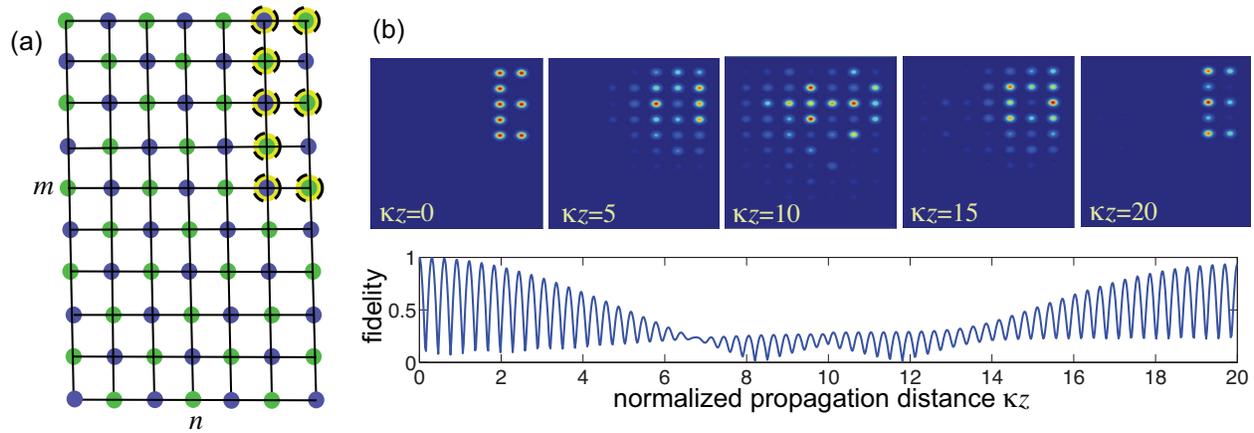}} \caption{ \small
(Color online) (a) Schematic of a bidimensional binary waveguide lattice made of $N=7 \times M=10$ waveguides. The lattice is initially excited by a single photon W state
that reproduces the letter E (the eight circled waveguides on the top right side of the lattice). (b) Numerically-computed evolution of the mean photon number $\langle \hat{a}^{\dag}_{n,m} \hat{a}_{n,m} \rangle$ (classical light intensity) for a few values of normalized propagation distance $\kappa z$ (upper panels) and of the fidelity $\mathcal{F}(z)$ (lower panel) for homogeneous coupling constants and for $\delta / \kappa=10$, $L=10 / \kappa$.}
\end{figure*}  
  While single photon state excitation basically reproduces the classical light propagation in the waveguide array, the fidelity decreases for photon number state excitation with a high photon number. In fact, indicating by $\mathcal{F}_1$ the fidelity corresponding to single photon excitation at waveguide $n_1$, i.e. $| \psi(0) \rangle = \hat{a}^{\dag}_{n_1} |0 \rangle$, it can be readily shown that the fidelity $\mathcal{F}_{N_0}$ corresponding to the excitation with the $N_0$ photon number (Fock) state $| \psi(0) \rangle = (1 / \sqrt{N_0 ! })  \hat{a}^{\dag N_0}_{n_1} |0 \rangle$ at guide $n_1$ is given by $\mathcal{F}_{N_0}=\mathcal{F}_{1}^{N_0}$, thus rapidly degrading as the photon number $N_0$ increases. A similar behavior is found when the array is excited with other non-classical states of light. Figure 3(a) shows, as an example, the behavior of the fidelity at the output plane  for the same homogeneous binary lattice of Fig.1 but when it is excited by by a NOON state  $|\psi(0)\rangle=(1/ \sqrt{2 N_0!}) ( \hat{a}^{\dag N_0}_{n_1} |0 \rangle + \exp(i \varphi)  \hat{a}^{\dag N_0}_{n_2} |0 \rangle)$ with $N_0=1$ (i.e. a Bell state) and $N_0=2$ photons at waveguides $n_1=1$ and $n_2=2$. The behavior of the fidelity for the NOON state excitation with $\varphi=0$ and for increasing photon number $N_0$ is shown in Fig.3(b). The result clearly indicates that, as the photon number increases, the revival to the initial state (Loschmidt echo) is rapidly degraded like for the Fock state excitation.  This result provides a kind of resolution of the Loschmidt 'paradox': for a 'macroscopic' (many-body) system , i.e. for a state with a large photon number, even a small deviation of $\hat{H}_2$ from $-\hat{H}_1$ makes it the backward dynamics  to largely deviate from the forward one, and thus reversibility of the dynamics unlikely. Note that, since interaction of photons with the environment is negligible, the decay of Loschmidt echo arises here because of imperfect time reversal  rather than from decoherence effects \cite{r1bis}, and can be fully controlled by varying the offset $\delta$.\\
 The previous analysis can be readily extended to a two-dimensional binary array, i.e. for a rectangular lattice comprising $(N \times M)$ waveguides with propagation constant offset $ (-1)^{n+m} \delta$ at lattice site $(n,m)$. As an example, Fig.4 shows the Loschmidt echo (self-imaging) dynamics in a binary array made of $N=7 \times M=10$ waveguides which is initially excited by a single photon in a W state \cite{r33} that reproduces the alphabetic letter E on the right top of the lattice [see Fig.4(a)], i.e. $|\psi(0) \rangle= (1/ \sqrt{8}) \sum_{ (n,m)} \hat{a}^{\dag}_{n,m} |0 \rangle$, where the sum is extend over the eight indices $(n,m)$ that identify the letter E [the circled sites in the lattice of Fig.4(a)]. Figure 4(b), upper panels, show the numerically-computed evolution of the mean photon number (classical light intensity) for a few values of the normalized propagation distance $\kappa z$, wheres the behavior of the fidelity $\mathcal{F}(z)$ is depicted in the lower panel. Note that the bidimensional lattice realizes an approximate self-imaging with a fidelity of $ \sim 0.94$ for the single photon state. Like for the one-dimensional case, the self-imaging turns out to be robust against disorder or imperfections in the lattice coupling constants. However, the fidelity rapidly degrades for excitation with  high photon number states.\\
 In conclusion, Loschmidt echo of photons propagating in binary waveguide lattices has been theoretically suggested. The revival does not require to reverse the full Hamiltonian of the system, since time reversal is approximately  obtained by exchanging the two sublattices of the array after some propagation distance, but not the sign of the coupling constants. Examples of Loschmidt echoes for single photon and NOON states have been given in one- and two-dimensional waveguide lattices. While high fidelity of time reversal can be observed for single photon excitation, that reproduces the fidelity of revival at the classical level, the fidelity is rapidly degraded when the lattice is excited by a state with a large number of photons, such as Fock or NOON states. Our results indicate that photonic transport in engineered waveguide lattices is an interesting platform to investigate decoherence-free time reversal dynamics at the quantum level \cite{r1bis} and Loschmidt echo decay  when the degrees of freedom of the system (i.e. the number of photons) is increased. Integrated quantum photonics, a rapidly emerging research area \cite{r19,r20,r33}, is expected to provide an experimentally accessible platform to test Loschmidt paradox and its resolution with photons.

\newpage


 {\bf References with full titles}\\
 \\
 \noindent
1. T. Gorin, T. Prosen, T.H. Seligman, and  M. Znidaric, {\it Dynamics of Loschmidt echoes and fidelity decay}, Phys.Rep. {\bf 435}, 33 (2006).\\
2. Ph. Jacquoda and C. Petitjean, {\it Decoherence, entanglement and irreversibility in quantum dynamical systems with few degrees of freedom}, Adv. Phys. {\bf 58}, 67 (2009).\\
3.  A. Goussev, R.A. Jalabert, H.M. Pastawski, and D.A. Wisniacki, {\it Loschmidt echo and time reversal in complex systems}, Phil. Trans. R. Soc A  {\bf 374}, 20150383 (2016).\\
4. E.L. Hahn,  {\it Spin echoes}, Phys. Rev. {\bf 80}, 580 (1950).\\
5. M. Fink, {\it Time reversed acoustics}, Phys. Today {\bf 50}, 34 (1997).\\
6. G. Lerosey, J. de Rosny, A. Tourin, A. Derode, G. Montaldo, and M. Fink, {\it Time Reversal of Electromagnetic Waves},
Phys. Rev. Lett. {\bf 92}, 193904 (2004).\\
7. A. Przadka, S. Feat, P. Petitjeans, V. Pagneux, A. Maurel, and M. Fink, {\it Time reversal of water waves}, Phys. Rev. Lett. 109, 064501 (2012).\\
8. J. Martin, B. Georgeot, and D. L. Shepelyansky, {\it Cooling by Time Reversal of Atomic Matter Waves}, Phys. Rev. Lett. {\bf 100}, 044106 (2008).\\
9. A. Alberti, G. Ferrari, V.V. Ivanov, M.L. Chiofalo, and G.M. Tino, {\it Atomic wave packets in amplitude-modulated vertical optical lattices}, New J. Phys. {\bf 12}, 065037  (2010).\\
10. F.M. Cucchetti, {\it Time reversal in an optical lattice}, J. Opt. Soc. Am. B {\bf 27}, A30 (2010).\\
11. A. Ullah and M.D. Hoogerland, {\it Experimental observation of Loschmidt time reversal of a quantum chaotic system}, Phys Rev E {\bf 83}, 
046218 (2011).\\
12. Z. Yaqoob,  D. Psaltis, M.S. Feld, and C. Yang, {\it Optical phase conjugation for turbidity suppression in biological samples},  Nature Photon. {\bf 2}, 110 (2008).\\ 
13. A.P. Mosk, A. Lagendijk, G. Lerosey, and M. Fink, {\it Controlling waves in space and time for imaging and focusing in complex media}, Nature Photon. {\bf 6}, 283 (2012).\\
14. J. Park, C. Park, K. R. Lee, Y.-H. Cho, and Y.K. Park, {\it Time-reversing a monochromatic subwavelength optical focus by optical phase conjugation of multiply-scattered light}, Sci. Rep. {\bf 7}, 41384 (2017).\\
15. G. Leuchs and M. Sondermann, {\it Time reversal symmetry in optics}, Physica Scr. {\bf 85}, 058101 (2012).\\
16. D. N. Christodoulides, F. Lederer, and Y. Silberberg, {\em Discretizing light behaviour in linear and nonlinear waveguide lattices}, Nature {\bf 424}, 817 (2003).\\
17. S. Longhi, {\it Quantum-optical analogies using photonic structures}, Laser \& Photon. Rev. {\bf 3}, 243 (2009).\\
18. I.L. Garanovich, S. Longhi, A.A. Sukhorukov, and Y.S. Kivshar, {\it Light propagation and localization in modulated photonic lattices and waveguides}, Phys. Rep. {\bf 518}, 1 (2012).\\
19. Y. Bromberg, Y. Lahini, R. Morandotti, and Y. Silberberg, {\it Quantum and Classical Correlations in Waveguide Lattices}, Phys. Rev. Lett. {\bf 102}, 253904 (2009).\\
20. A. Peruzzo, M. Lobino, J.C.F. Matthews, N. Matsuda, A. Politi, K. Poulios, X.-Qi Zhou, Y. Lahini, N. Ismail, K. W\"orhoff, Y. Bromberg, Y. Silberberg, M.G. Thompson, and J.L. O'Brien, {\it Quantum walks of correlated particles}, Science, {\bf 329}, 1500 (2010).\\
21. T. Meany, M. Gr\"afe, R. Heilmann, A. Perez-Leija, S. Gross, M.J. Steel, M.J. Withford,
and A. Szameit, {\it Laser written circuits for quantum photonics}, Laser \& Photon. Rev. {\bf 9}, 363 (2015).\\
22. R. Morandotti, U. Peschel, J. S. Aitchison, H. S. Eisenberg, and Y. Silberberg, {\it Experimental observation of linear and nonlinear optical Bloch oscillations}, Phys. Rev. Lett. {\bf 83}, 4756 (1999).\\
23. T. Pertsch, P. Dannberg, W. Elflein, A. Brauer, and F. Lederer, {\it Optical Bloch Oscillations in Temperature Tuned Waveguide Arrays}, Phys. Rev. Lett. {\bf 83}, 4752 (1999).\\
24. F. Dreisow, A. Szameit, M. Heinrich, T. Pertsch, S. Nolte, A. T\"unnermann, and S. Longhi, {\it Bloch-Zener oscillations in binary superlattices},
Phys. Rev. Lett. {\bf 102}, 076802 (2009).\\
25. T. Schwartz, G. Bartal, S. Fishman, and M. Segev, {\it Transport and Anderson localization in disordered two-dimensional photonic lattices}, Nature {\bf 446}, 52 (2007).\\
26. Y. Lahini, A. Avidan, F. Pozzi, M. Sorel, R. Morandotti, D. N. Christodoulides, and Y. Silberberg, {\it Anderson Localization and Nonlinearity in One-Dimensional Disordered Photonic Lattices}, Phys. Rev. Lett. {\bf 100}, 013906 (2008).\\
27. F. Dreisow, A. Szameit, M. Heinrich, T Pertsch, S. Nolte, A. T\"unnermann, and S. Longhi, {\it Decay control via discrete-to-continuum coupling modulation in an optical waveguide system}, Phys. Rev. Lett. {\bf 101}, 143602 (2008).\\ 
28. S. Longhi, M. Marangoni, M. Lobino, R. Ramponi, P. Laporta, E Cianci, and V. Foglietti,
{\it Observation of dynamic localization in periodically curved waveguide arrays }, Phys. Rev. Lett. {\bf 96}, 243901 (2006).\\
29. L. Sansoni, F. Sciarrino, G. Vallone, P. Mataloni, A. Crespi, R. Ramponi, and R. Osellame, {\it Two-particle bosonic-fermionic quantum walk via integrated photonics}, Phys. Rev. Lett. {\bf 108}, 010502 (2012).\\
30. G. Di Giuseppe, L. Martin, A. Perez-Leija, R. Keil, F. Dreisow, S. Nolte, A. Szameit, A. F. Abouraddy, D. N. Christodoulides, and B. E. A. Saleh, {\it Einstein-Podolsky-Rosen Spatial Entanglement in Ordered and Anderson Photonic Lattices}, Phys. Rev. Lett. {\bf 110}, 150503 (2013).\\
31. H.S. Eisenberg, Y. Silberberg, R. Morandotti, and J.S. Aitchison, {\it Diffraction management},
Phys. Rev. Lett. {\bf 85}, 1863 (2000).\\
32. S. Longhi, {\it Image reconstruction in segmented waveguide arrays}, Opt. Lett. {\bf 33}, 473 (2008).\\
33. A. Szameit, F. Dreisow, M. Heinrich, T. Pertsch, S. Nolte, A.
T\"unnermann, E. Suran, F. Louradour, A. Barthelemy, and S.
Longhi, {\it Image reconstruction in segmented femtosecond laser-written waveguide arrays}, Appl. Phys. Lett. {\bf 93}, 181109 (2008).\\
34. R. Keil, Y. Lahini, Y. Shechtman, M. Heinrich, R. Pugatch, F. Dreisow,
A. T\"unnermann, S. Nolte, and A. Szameit, {\it Perfect imaging through a disordered waveguide lattice}, Opt. Lett. {\bf 37}, 809 (2012).\\
35. R. Morandotti, D. Mandelik, Y. Silberberg, J.S. Aitchison, M. Sorel, D.N. Christodoulides, A.A. Sukhorukov, and Y.S. Kivshar, 
{\it Observation  of discrete gap solitons in binary waveguide arrays}, Opt. Lett. {\bf 29}, 2890 (2006).\\
36. S. F\"olling, S. Trotzky, P. Cheinet, M. Feld, R. Saers, A. Widera, T. M\"uller, and  I. Bloch, 
{\it Direct observation of second-order atom tunnelling}, Nature {\bf 448}, 1029 (2007).\\
37. G. Corrielli,  A. Crespi, G. Della Valle, S. Longhi, and R. Osellame, {\it Fractional Bloch oscillations in photonic lattices}, Nature Commun. {\bf 4}, 1555 (2013).\\
38. M. Gr\"afe,	R. Heilmann, A. Perez-Leija, R. Keil,	F. Dreisow, M. Heinrich, H. Moya-Cessa, S. Nolte, D.N. Christodoulides, and A. Szameit,
 {\it  On-chip generation of high-order single-photon W-states}, Nature Photon. {\bf 8}, 791 (2014).

\end{document}